# Analyzing DevOps Practices Through Merge Request Data: A Case Study in Networking Software Company

Samah Kansab[a], Matthieu Hanania[a], Francis Bordeleau[a], Ali Tizghadam[b]

[a]*École de technologie supérieure (ÉTS), Montreal, Canada*
[b]*TELUS, Toronto, Canada*

**Abstract**

DevOps integrates collaboration, automation, and continuous improvement, enhancing agility, reducing time to market, and ensuring consistent software releases. A key component of this process is GitLab's Merge Request (MR) mechanism, which streamlines code submission and review. Studies have extensively analyzed MR data and similar mechanisms like GitHub pull requests and Gerrit Code Review, focusing on metrics such as review completion time and time to first comment. However, MR data also reflects broader aspects, including collaboration patterns, productivity, and process optimization.

This study examines 26.7k MRs from four teams across 116 projects of a networking software company to analyze DevOps processes. We first assess the impact of external factors like COVID-19 and internal changes such as migration to OpenShift. Findings show increased effort and longer MR review times during the pandemic, with stable productivity and a lasting shift to out-of-hours work, reaching 70% of weekly activities. The transition to OpenShift was successful, with stabilized metrics over time. Additionally, we identify prioritization patterns in branch management, particularly in stable branches for new releases, underscoring the importance of workflow efficiency. In code review, while bots accelerate review initiation, human reviewers remain crucial in reducing review completion time. Other factors, such as commit count and reviewer experience, also influence review efficiency.

This research provides actionable insights for practitioners, demonstrating how MR data can enhance productivity, effort analysis, and overall efficiency in DevOps.

# 1. Introduction

DevOps represents an integrated approach that emphasizes collaboration, automation, and continuous improvement throughout the software development lifecycle [8, 20]. By integrating DevOps practices, organizations can significantly enhance their agility, shorten the time to market, increase automation, and ensure more reliable and consistent software releases [3]. DevOps integrates the efforts of development and operations teams, facilitates continuous integration, continuous delivery, and continuous deployment, and fosters a culture of collaboration and shared responsibility to address common bottlenecks and inefficiencies in traditional software development practices [10]. The DevOps process is structured into distinct phases, each crucial for maintaining a smooth and efficient workflow, which are: development, testing, building and monitoring. Each phase necessitates specific checks and validations before progressing to the next phase. By leveraging DevOps within the framework of SPI, organizations can achieve a streamlined and effective software development lifecycle, ultimately leading to higher productivity, better software quality, and reduced development costs and time [17, 19, 21, 26].

To realize the DevOps process, multiple tools and technologies are employed at each stage, ensuring seamless execution and continuous improvement throughout the software development process. For instance, in the development stage, version control systems like Git are used, while in the testing stage, continuous integration (CI) servers such as GitHub Actions or GitLab CI are common. Additionally, there are different mechanisms that help streamline the workflow, such as GitLab's merge request and Git hooks.

GitLab's Merge Request (MR) mechanism serves as a central element within the DevOps pipeline, streamlining the submission, review, and integration of new code changes into the main code base (from source to target branch) [1]. Throughout this process, collaborators actively participate in discussions, offering insights and improvements to the code under review. In addition, the MR encapsulates various aspects of the project and team dynamics. Analyzing this data provides a rich perspective on multiple dimensions of the software system. It facilitates productivity analysis by quantifying effort and contributions, identifies workflow bottlenecks to optimize code updates, ensures quality of work through thorough reviews, and improves collaboration patterns by identifying key contributors. It also provides insight into how the development process is impacted by various factors, such as the impact of new technology integration on release velocity.

While MR data have been used in the literature to analyze different aspects of the code review process [7, 4, 16, 9], the goal of this paper is to examine various aspects of DevOps processes using GitLab MR data to identify and improve areas affected by different changes in the process. To do so, we analyze 26,7k MRs data from four teams of our industrial partner, a networking software solution company that deploys 5G and cloud edge fabric technologies for telco, data center, and cloud service provider customers using a DevOps approach for efficient software development and deployment. These main aspects are addressed in this study:

- **Impact of Environmental or Process Changes:** Changes within or outside the company can significantly affect team dynamics and performance. Analyzing MR data reveals how these changes impact productivity, collaboration, and effort, helping us address two key questions:

  - **Environmental Changes:** we examine the case of the COVID-19 pandemic and answer the research question: ***How have environmental changes, specifically the shift from office-based work to remote work during the COVID-19 pandemic, impacted workplace dynamics and productivity?*** Analyzing MR data shows increased effort and longer times to complete code review during COVID-19, with more activity outside regular hours and weekends, which persist after the pandemic, indicating greater flexibility. Despite these changes, productivity remained stable, demonstrating team adaptability.

  - **Process Changes:** We explore the impact of migrating to OpenShift technology. Our research question is: ***Has there been any impact of the process changes, specifically the migration to OpenShift technology, on different aspect of the organization?*** During the migration, we observe unstable times to complete code review and smaller MRs on OpenShift due to gradual technology integration and code changes. By the final sprints of the migration, times to complete code review and MR sizes stabilized, showing increased developer competency with OpenShift.

- **Branch Information Analysis:** We compare MR data across different branch types (master, stable, other) to understand how teams incorporate changes and deploy software, answering: ***How are different types of GitLab branches used and managed in an industrial DevOps environment?*** We found that MRs on stable branches are prioritized, completed faster, and receive quicker responses, likely due to their connection to release production and bug management.

- **Code Review Process Analysis:** We analyze the timing and nature of initial feedback in code reviews to understand its impact on MR efficiency in DevOps. Our research question is: ***How do the timing and type of initial feedback in code reviews impact MR efficiency in DevOps, and what roles do bots and human interactions play?*** We found that bot-generated first comments dominate initial MR interactions, significantly speeding up the initiation of the review process, with median response times ranging from 4.46 to 23.78 minutes. However, it is the human comments that have the most significant impact on reducing time to complete code review, explaining over 93% of the variance in MR completion time. This finding highlights the critical role of timely human feedback in driving efficient code reviews, while also acknowledging the importance of bots in initiating the process. Other factors, such as the number of commits and the experience of the reviewers, were also found to play a significant role in determining the overall efficiency of the code review process.

---

[1]https://docs.gitlab.com/ee/user/project/merge_requests/

The main contribution of this study lies in highlighting the broader significance of MR data, extending its analysis beyond traditional code review to provide insights into various aspects of the DevOps process. By examining MR data, we gain a deeper understanding of not only the quality of code reviews but also how teams collaborate and how external factors, such as technological changes or unexpected events, influence the development workflow. Additionally, this study offers a valuable analysis of the factors impacting review completion times and the timing of the first comment in an industrial context. Our research provides practical guidance for practitioners on leveraging code review data, particularly MR data, to enhance their DevOps processes through more in-depth analysis and data-driven decision-making.

The paper proceeds with related work in Section 2, followed by the methodology in Section 3. Section 4 addresses the research questions, concluding with discussions on threats to validity in Section 5 and the paper's conclusion in Section 6.

## 2. Background and Related Work

GitLab's MR mechanism is a structured process, an essential aspect of DevOps, that facilitates the proposal and incorporation of new code changes into the main code base. This process begins when an MR is created to integrate the proposed changes into the target branch. Throughout the MR lifecycle, various statuses are assigned, including "Open" for review, "Closed" if changes are considered unnecessary, and "Merged" once the changes have been successfully integrated. Collaboration among contributors, including the author, committers, reviewers, commenters, and integrators, is paramount in assessing code quality and making decisions about proposed changes.

While GitLab's MR mechanism is a cornerstone of code review within the DevOps framework, similar mechanisms exist on other platforms, such as pull requests on GitHub and code review mechanisms on Gerrit. Pull requests are very close to MRs, with similar processes and underlying concepts, though on different platforms. All of these processes serve as an essential step in ensuring that code meets quality standards before integration into the main codebase [27], and has been extensively studied in the literature.

Researchers have extensively studied data from code review mechanisms to analyze and explain the review process. For example, Chouchen et al. [4] and Maddila et al. [16] proposed models for predicting review completion time and identifying factors affecting it. Hasan et al. [7] observed that shorter time-to-universal-first-response correlates with shorter pull request lifetimes, except for bot-first pull requests. Similarly, Bosu and Carver [2] found that prominent developers receive faster initial feedback, resulting in quicker overall reviews. Jiang, Adams, and German [11] demonstrated that the number of reviewers influences review duration, while Thongtanunam et al. [25] explored the code patch characteristics that attract more reviewers. Baysal et al. [1] showed that both technical and non-technical factors affect the time it takes to complete reviews. Fan et al. [5] and Islam et al. [9] used machine learning techniques to predict merge likelihood and related factors. Li et al. [15] analyzed the resource consumption caused by duplicate pull requests and how selection criteria are applied. Additionally, Khatoonabadi et al. [12] investigated factors contributing to abandoned pull requests, revealing that these often involve less experienced contributors, increased complexity, and longer review processes.

Researchers have also used review data to examine coding-related factors. For instance, Nejati, Alfadel, and McIntosh [18] found that build specification changes in the Qt[2] and Eclipse[3] projects are discussed less frequently during reviews than production and test code, but are more likely to reveal defects, especially those related to evolvability and dependencies. Spadini et al. [22] showed that production code tends to receive more attention than test code during reviews, though both are significant. Likewise, Thongtanunam et al. [24] discovered that risky or defective files receive less thorough reviews, with feedback more often focused on improving evolvability, such as documentation, rather than functional issues.

In addition to studying code review practices and coding factors, the data obtained through this process provides valuable insights for deeper analysis of the DevOps workflow. In this paper, our goal is to study various aspects of the DevOps process using MR data in an industrial context, including the impact of process or environment changes, branch analysis, and finally, code review analysis, similar to previous studies. Our goal is to improve the DevOps process by drawing conclusions to ensure its smooth operation and resilience to changes.

## 3. Methodology

The goal of this section is to discuss the steps that we followed to collect the metrics to investigate different aspects of DevOps processes that can be analyzed using GitLab MR data. Below, we explain the broad outlines of our process, shown in Figure 1, leaving the details to the research questions.

### 3.1. Data collection

In this step, we use a GitLab extraction tool, developed by Legault [14] as a base to develop our data extraction tool using Gitlab API[4], to extract the MR data using the GitLab API. We collect in this study the data from 4 teams with many projects, as shown in 2. The **Control Plane (CP)** and **Management Plane (MP)** primarily utilize Golang for managing network resources, orchestrating traffic routing, and overseeing device configuration, monitoring, and provisioning. The **Platform**

---

[2]https://contribute.qt-project.org/
[3]https://eclipse.dev/eclipse/
[4]https://docs.gitlab.com/ee/api/merge_requests.html

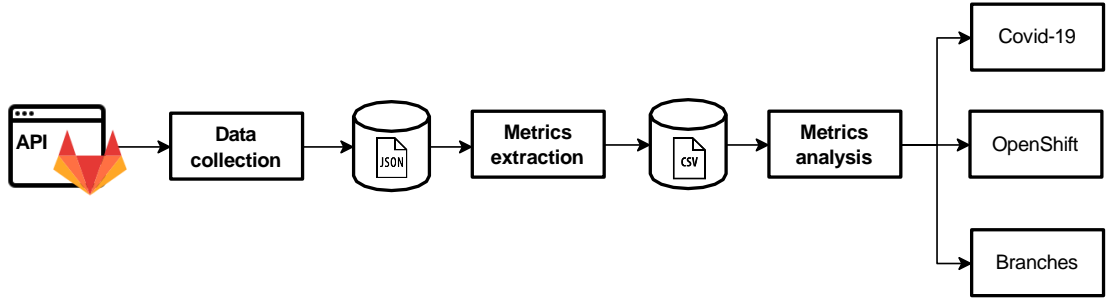

Figure 1: The architecture of our study

**(PF)** group specializes in low-level software components like kernel-level development and system libraries, while the **Data Plane (DP)** group focuses on packet forwarding and processing, utilizing different technologies, such as P4 for networking devices and relying on hardware integration testing, alongside the **FPGA** team. We collected data from January 1, 2019, to June 30, 2023. The raw data gathered includes comprehensive information on MRs such as ID, creation date, and end date. Additionally, we collected details on commits related to each MR, discussions and notes, and changes made to each file. This extensive data collection provides the flexibility needed for thorough analysis.

*3.2. Metrics Extraction*

Once the raw data is collected, we proceed to calculate various metrics crucial for analyzing different aspects of the DevOps process using MR data. For example, the time to complete code review, which is not directly available from the GitLab API, needs to be calculated as the time between the creation and closure of the MR. In total, 24 metrics, as shown in Table 1, are collected across three dimensions: the MR dimension, which includes metrics related to each MR; the project dimension, which consists of project information; and the developer dimension, which includes metrics related to team members. These metrics provide a comprehensive view of the DevOps process, facilitating detailed analysis and insights.

*3.3. Metrics Analysis*

We conduct different types of analysis for each research question (RQ). To do this, we filter the data by two time dimensions: by week or by sprint, with a sprint representing the company's three-week development cycle. Once the data is filtered, we use mathematical values such as mean, median, minimum, and maximum to draw conclusions and compare different aspects. To ensure our analysis is coherent, we fixed several aspects to be analyzed, such as productivity (measured by the number of MRs created and closed each week) and time to complete code review (to measure effort). Each aspect and method is detailed in the approach of each RQ.

Table 1: List of Mtrics Used in This Study

| Metric | Description |
|---|---|
| Creation Week ‡ | The week during which the MR was created. |
| Creation Sprint † | The sprint during which the MR was created. |
| End Week ‡ | The week during which the MR was closed. |
| Time to complete code review (Lead Time) ‡ † ∗ ⋆ | The duration between the creation and closure of the MR. |
| Creation Hour ‡ | The hour when the MR was initially opened. |
| Closure Hour ‡ | The hour when the MR was closed. |
| Is ORH ‡ | A flag that indicates whether the MR includes activities occurring outside of regular working hours. |
| MR Size † ∗ | The amount of lines of code added + deleted related to the MR. |
| Additions † | The amount of added lines of code related to the MR. |
| Deletions † | The amount of deleted lines of code related to the MR. |
| Source Branch ∗ | The branch to which the code is added during the MR process. |
| Target Branch ∗ | The branch to which the code is merged upon the closure of the MR. |
| Time to First Comment ∗ ⋆ | Time taken to receive the first comment from the creation of the MR. |
| #Notes ∗ | Number of discussion messages made during the review of the MR. |
| Description Length ⋆ | The length of the description provided for the MR. |
| Is First MR ⋆ | Indicates whether this is the first MR submitted for a given project. |
| Is Hash-tag ⋆ | Indicates whether the "#" tag is present in the MR's title or description. |
| Is At-Tag Presence ⋆ | Indicates whether the "@" tag is present in the MR's title or description. |
| #Commits ⋆ | The number of commits made at the time the MR was opened. |
| Project Age ⋆ | The number of months from the project's creation to the time of the MR submission. |
| #Opened MRs ⋆ | The number of MRs that are open for a given project at the time of analysis. |
| #Merges ⋆ | The number of prior merges in the project. |
| #Previous MRs ⋆ | The number of all previous MRs that have been previously created for the project. |
| Author as Reviewer ⋆ | Indicates whether the MR author is also the reviewer. |

‡ **Corresponds to RQ1,** † **Corresponds to RQ2,** ∗ **Corresponds to RQ3,** ⋆ **Corresponds to RQ4**

Table 2: Project teams of our industrial partner

| Group | Description | #MR | #Projects |
|---|---|---|---|
| Management Plane (MP) | Manage the communication with Fabric | 6,344k | 20 |
| Control Plane (CP) | Orchestrate paths for packets and frames | 8,396k | 31 |
| Data Plane (DP) | Forward packet and frames between interfaces | 7,416k | 16 |
| FPGA | designs reprogrammable integrated circuits | 735 | 19 |
| Platform (PF) | Manage low-level platforms | 4,004k | 30 |

## 4. Results

**RQ1: How have environmental changes, specifically the shift from office-based work to remote work during the covid-19 pandemic, impacted workplace dynamics and productivity?**

*Motivation*

The onset of the COVID-19 pandemic in Canada on March 16, 2020, necessitated an immediate shift from office-based to remote work practices, requiring a fundamental rethinking of established work habits. Previously, daily interactions in the office facilitated discussions about development activities and code reviews over coffee. With the transition to remote work, however, reliance on corporate technologies and communication mechanisms became imperative. In this study, we seek to investigate the impact of the pandemic on developer activities by analyzing MR data. Our overarching goal is to explore how external changes, such as the pandemic, affect the DevOps process and reflect the organization's flexibility in adapting to abrupt circumstances.

*Approach*

To investigate the impact of Covid-19 on various aspects of the DevOps process, we conducted a comparative analysis before and after the pandemic, focusing on the following key issues:

- **Productivity:** to have clearer vision, we conduct this analysis spanning the first 22 weeks of 2020. This period includes 11 weeks before and 11 weeks after the start of the pandemic-induced lockdown, which began at week 12. We assessed team productivity by examining the number of MRs created and ended each week before and after the Covid-19 outbreak, including transition week 12. Tracking the number of MRs over time helps us see how fast development tasks are being merged and completed. More MRs created could mean that more work is being done, indicating higher productivity. But if there's a big decrease or change in the number of MRs completed, it could mean there are problems or delays in the development process. This analysis helps teams identify trends and issues in their DevOps workflow.

- **Efforts:** In assessing the impact of Covid-19 on developer effort, we study time to complete code review on the 22 weeks, which represents the time it takes to complete a code review. This involved analyzing the time taken to complete code reviews both before and after the pandemic, as well as the number of successfully merged MRs. A longer time to complete code review suggests that collaborators spent more time reviewing, providing feedback, and updating the code, indicating an increased effort to ensure code quality and accuracy. Alternatively, it could indicate delays in the code review process, resulting in a slower overall timeline. Conversely, shorter review times may indicate less effort (a simpler MR) in the review process, or they may reflect the prioritization of the work, necessitating a quicker completion. Therefore, as supported by previous research [4, 16, 1, 11], we consider the time to complete code review as a proxy to analyze the effort to complete the review process.

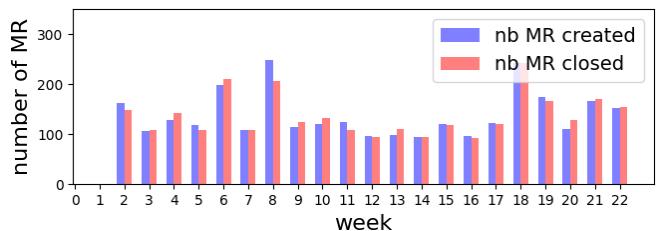

Figure 2: Number of MR created and ended per week from 1 to 22 of 2020

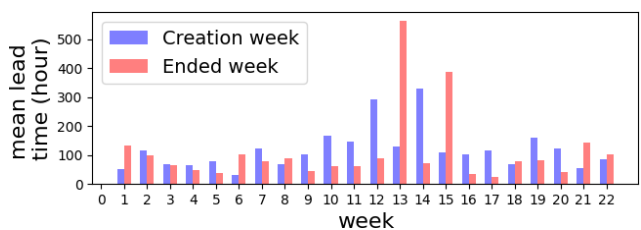

Figure 3: Mean time to complete code review by week for MR opening and closing from week 2 to 22 of 2020

- **Temporal Autonomy:** With the transition to remote work, individuals have gained greater flexibility in their work schedules, allowing them to work during nights, weekends, or holidays. In this phase, we examine the impact of Covid-19 on developers' temporal autonomy by analyzing their activity on MRs before 8 a.m. and after 5 p.m. This includes actions such as creating or ending MRs, reviewing, and committing. For each week, we calculate the percentage of these activities occurring outside regular working hours compared to the total number of activities throughout the entire day. This analysis includes all data to study the persistence of the pandemic's impact on developers' habits years after the pandemic.

*Results*

**Despite the onset of the Covid-19 pandemic, the productivity of the teams of the studied company, as measured by the weekly number of MRs created and closed, remained relatively stable.** Figure 2 illustrates this trend, showing that there was no significant change in the volume of MRs despite the onset of the pandemic. Specifically, there were 125 MRs created in week 11 (the last week of work in the office), which decreased slightly to 96 in week 12, a decrease of 23.2%. Similarly, the number of MRs closed decreased slightly from 109 in week 11 to 93 in week 12, a decrease of 14.68%. Regarding week 13 (the first week after the restriction decision), we observe that the number of MRs created is the same as in week 12. We also notice that the number of MRs closed exceeds the number of MRs created in this week. However, we observe this pattern (number of closed MRs exceeding the number of created MRs) in other weeks prior to Covid, such as weeks 4, 9, and 10, indicating that this is not solely the impact of Covid. These results suggest that, contrary to expectations, the Covid-19 pandemic did not have a significant impact on the productivity levels of the teams when measure based on their MR activity.

*Note: The first week of the year 2020 is empty. Because on this week, no developers have created (open) or end (close) any MR while it was January holidays.*

**Contrary to our initial findings, we observe a significant increase in time to complete code review with the onset of COVID-19. This indicates either greater effort or more**

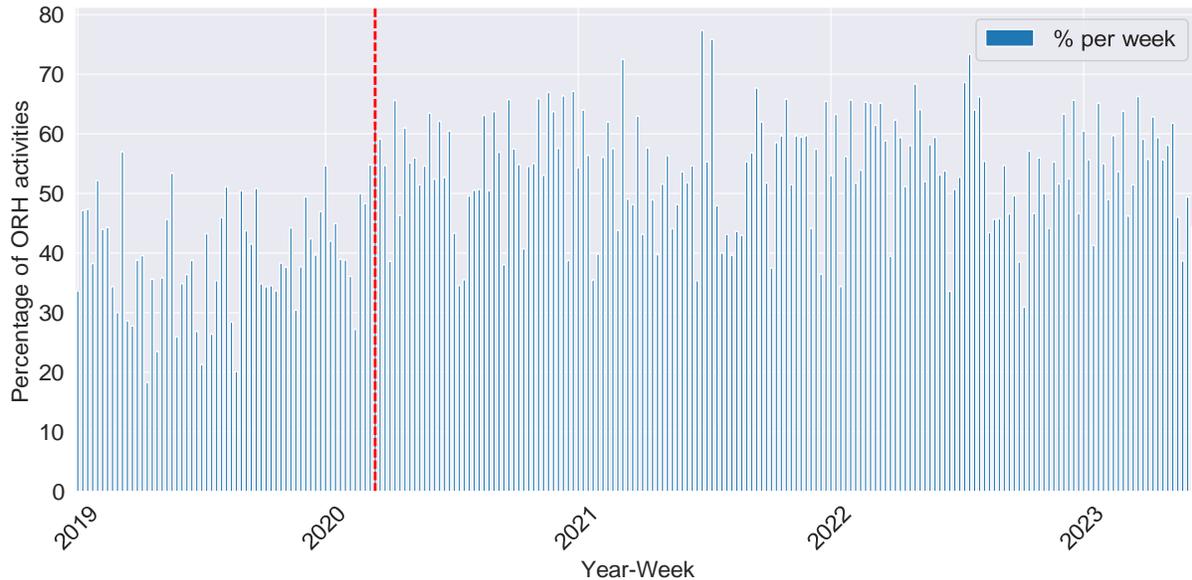

Figure 4: The distribution of the percentage of out of regular working hours activities by week before and after Covid-19

**delays during the review process by team members during this period.**. As shown in Figure 3, the mean time to complete code review for MRs created in week 10 jump from about 3.47 days to about 9.95 and 7.92 days in weeks 11 and 12, respectively, coinciding with the transition to remote work. This trend is similar to that observed in week 2 (after the January holidays), confirming the impact of Covid-19 on the time to complete code review of MRs created in weeks 11 and 12. Regarding the average time to complete code review of closed MRs, those closed in week 13 have an average of about 16.44 days, which is the greater average of the year. These values may explain the two spikes in MR creation observed in weeks 11 and 12. These results help explain the first finding, since it takes more effort to create and close the same number of MRs for the same level of productivity.

**As shown in table 3, activities on MRs outside regular working hours (8 a.m. to 5 p.m.) and on weekends nearly doubled after the pandemic, with up to 70.48% of activities occurring outside regular working hours.** As shown in Figure 4, there was a notable variation in the percentage of activities done before and after Covid-19. Table 3 shows that the mean and median percentages of weekly activities outside regular working hours were 38.50% and 39.45%, respectively, before the pandemic, which increased to 54.56% and 55.91% after the pandemic. Similarly, weekend activities increased significantly, with the median rising from 3.37% to 7.14%, as shown in Figure 5. This pattern clearly illustrates the impact of Covid-19 on developers' temporal

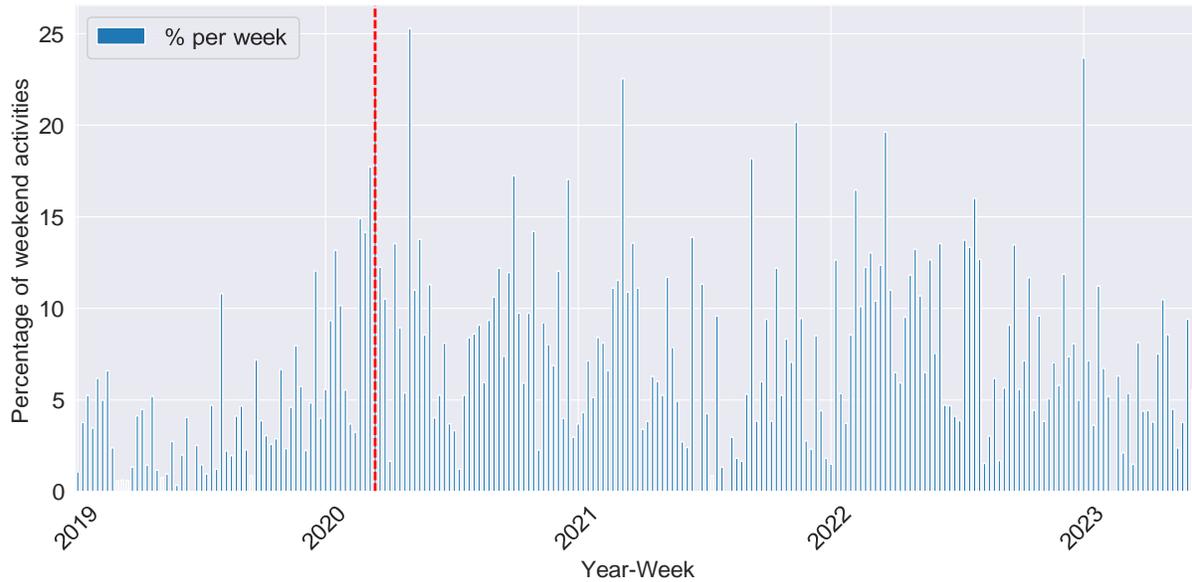

Figure 5: The distribution of the percentage of weekend activities by week before and after Covid-19

Table 3: Out of Regular Hours (ORH) and weekend statistics activities before and after COVID-19

| Period | #MRs | Pattern | Mean % | Median % | Min % | Max % |
|---|---|---|---|---|---|---|
| before covid | 8.8k | ORH | 38.50 | 39.45 | 10.52 | 58.53 |
|  |  | Weekends | 4.49 | 3.37 | 0 | 17.74 |
| after covid | 17.94k | ORH | 54.56 | 55.91 | 34.72 | 70.48 |
|  |  | Weekends | 7.95 | 7.14 | 0 | 25.28 |

autonomy. Working remotely allowed developers more flexibility, not being restricted to the 8 a.m. to 5 p.m. schedule, a trend that has persisted even after the pandemic.

> *RQ1 summary:* During COVID-19, Our industrial partner's teams showed increased effort with longer times to complete code review and increased activity outside of regular business hours. Notably, this after-hours activity persisted after the pandemic, indicating a lasting impact. Despite these changes, productivity levels remained stable, demonstrating the team's adaptability and efficiency in the face of abrupt change.

**RQ2: Has there been any impact of the process changes, specifically the migration to openshift technology, on different aspect of the organization?**

*Motivation*

In addition to environmental changes, process changes, especially the integration of new technologies, can impact the DevOps workflow. Developers need to learn new skills, refactor existing code if necessary, and ensure the proper functioning of updated software, which can introduce numerous bugs and require more effort. A concrete example is the migration to OpenShift technology in the studied company. They wanted to ensure a smooth transition by separating OpenShift-related MRs from non-related areas. This migration process led to the creation of 2.0 and spanned from May 1, 2020 to January 31, 2021. In this research question, we aim to analyze the impact of this migration on the overall process by comparing the MRs created for OpenShift and the other MRs that were not related to OpenShift.

*Approach*

To study the impact of the migration to OpenShift, we categorize MRs (MRs) into two groups: those created specifically for the OpenShift migration, targeting a dedicated branch, and other MRs created during the same period of the migration but not related to OpenShift. Considering company's three-week sprint cycle, we study three aspects:

- **Efforts:** We examine the time to complete code review of MRs, which reflects the effort spent by teams to integrate OpenShift, compared to the effort of completing the other MR group.

- **Change Magnitude:** We assess the MR size, represented by the number of added and deleted lines, to indicate the magnitude of changes introduced to the code. We also compare the number of lines added versus deleted to determine whether developers are primarily adding new code or correcting existing code (by deleting and replacing lines of code).

*Results*

**Our results show an inverse relationship between the time to complete code review of MRs for OpenShift and other MRs during the first eight sprints,** as shown in Figure 6. When significant effort was spent on OpenShift integration, the time to complete code review for other MRs increased, suggesting that the focus on OpenShift integration was causing delays in other areas. Beginning in the eighth sprint, the times to complete code review for both categories began to converge and vary similarly, eventually converging in the final sprint. This trend suggests that developers initially put extra effort into learning and integrating OpenShift, but as they became more familiar with the technology, the time required for OpenShift MRs converged with the time to complete code review required for other MRs.

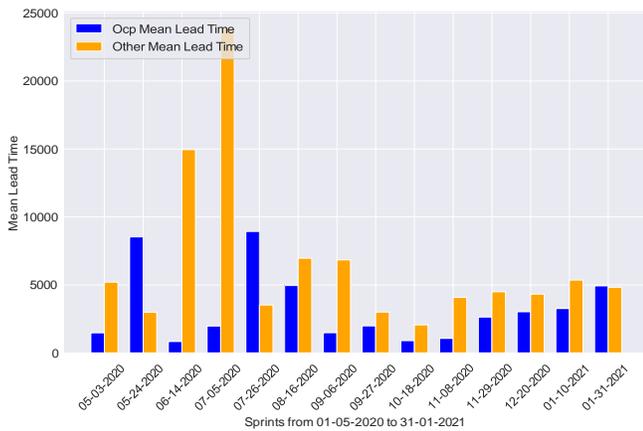 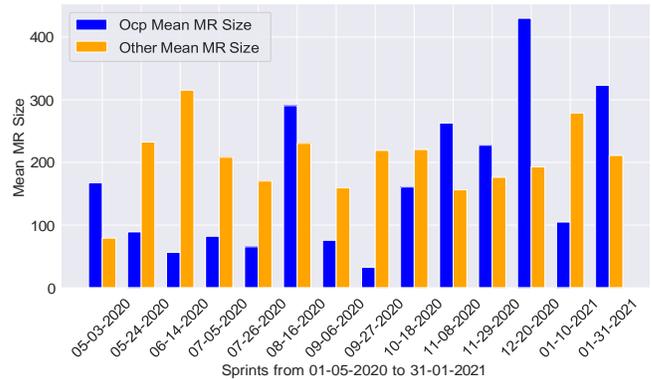

Figure 6: Mean time to complete code review by sprint for OpenShift MRs and other MRs in the migration period

Figure 7: Mean MR size by sprint for OpenShift MRs and other MRs in the migration period

**The MR sizes for OpenShift initially were smaller than other MRs but increased as developers gained expertise.** As shown in Figure 7, until Sprint 8, we observe that MR sizes for OpenShift are smaller compared to others, except for Sprint 6, indicating that developers were slowly integrating new changes into the software. This gradual integration resulted in smaller MR sizes for OpenShift compared to other projects. Starting in Sprint 8, MR sizes for OpenShift begin to increase, becoming larger than other MRs except for Sprint 13. This trend suggests that developers have become more skilled at integrating and reviewing larger chunks of code related to OpenShift.

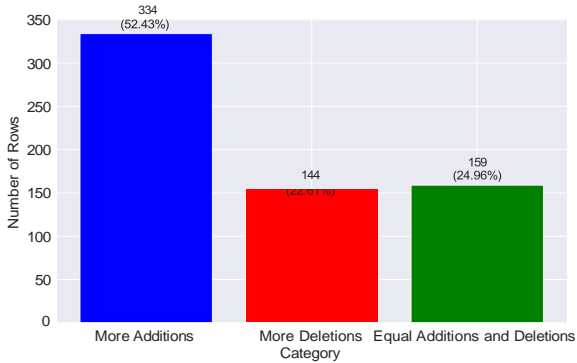 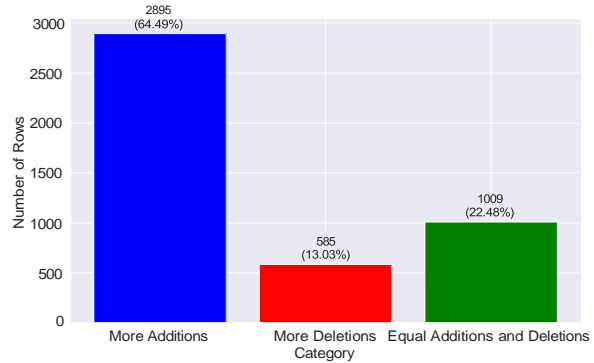

(a) Openshift MRs

(b) Other MRs

Figure 8: Comparison of Additions and Deletions per MR in Ocp and other MRs

**We observe distinct patterns in the nature of MRs for OpenShift and other MRs.** For other MRs, 86.97% involve more additions (64.49%) or equal additions and deletions (22.48%), indicating that these MRs are primarily for adding new code or modifying existing code during the review process. Only 13.03% of other MRs involve more deletions than additions. In contrast, 22.61% of OpenShift MRs involve more deletions than additions, indicating a focus on removing obsolete code. In addition, 52.43% of OpenShift MRs involve more additions, reflecting the integration of new technology and the introduction of new code. In addition, 24.88% of OpenShift MRs involve equal numbers of additions and deletions, indicating significant rework during the code review process. Such dynamics highlight that modifying the existed code (in 47.57% of OpenShift MRs) can impact code stability, that is considered as an indicator to the product (software) stability [23].

> *RQ2 summary:* The migration to OpenShift initially resulted in unstable times to complete code review and smaller MRs due to the gradual incorporation of new technology and required code changes during the review process. In the final sprints, times to complete code review and MR sizes stabilized, indicating the developers' increased skill with OpenShift.

## RQ3: How are different types of gitlab branches used and managed in an industrial devops environment?

*Motivation*

The branch policy in DevOps plays a critical role in ensuring smooth and efficient software development process. It allows teams to work independently on different features or fixes without disrupting the main codebase by isolating changes. By exploring how branches are used, teams gain insight into their organization and efficiency. They can monitor the progress of changes from starting to completion and deployment, and identify and address any problems or slowdowns along the way. For example, a rework made on the stable or final branch indicate a late-discovered bug. In addition, analyzing branch usage patterns reveals how the team operates, such as release frequency. MR data provides valuable insight into branch policy, as it specifies the source and target branches for each change, allowing a comprehensive study of these aspects. In this research question, we aim to investigate and understand the difference between different branch types and how they are managed using MR data.

*Approach*

To analyze the used branch categories using MR data at the studied company, we begin by understanding the company's organization, which operates on a yearly basis divided into seventeen three-week sprints, from Monday to Sunday. GitLab's branch structure remains consistent over the years, with the following categorization:

- **Main branch (Master):** serves as the primary development branch that remains consistent over time, acting as the base for daily development activities and the integration of new features for the latest and greatest product version.

- **Stable branches:** are dedicated to fix bugs and stabilize code quality for a planned release. Stable branches are branched out upon feature complete milestone of a planned release to distinguish between future releases development on master/main and the planned release bug fixes. Different planned releases and their corresponding bug fixes and patch releases are published and released from these stable branches.

- **Temporary branches:** are created for short-term development tasks or special initiatives, such as feature experiments or significant transitions like the migration to OpenShift, these branches are intended for limited use and are merged or discarded once the task is completed.

*Note: As shown in Figure 9, the number of MRs on the main branch (master) (23K) is always larger than the number of MRs on other branches (1.4K) and stable branches (1.9K), as the daily work is done on the master branch.*

Analyzing MRs (MRs):

- **Effort and Change Magnitude:** analyze the effort and the size of MRs similarly to the previous RQs.

- **Collaboration Aspects:** analyze the attention received by developers on MRs on different branches, looking at the time to receive the first comment on the MR and the number of discussion notes (messages) required to complete the review process.

By structuring our methodology in this manner, we ensure a comprehensive analysis of the branches, focusing on both the effort and the collaboration required for each category.

*Results*

**Although the MR size distribution of stable and master branches are very similar, we observe that the review process on the stable branch are completed faster than those on the master and other branches.** As shown in Figure 11, the MR size on the other branch category is larger than on the stable and master branches, which have almost the same distribution. Despite the larger size of MRs on the other and master branches, Figure 10 demonstrates that MRs on the stable branch are merged or closed more quickly. Table 4 further highlights this point, showing that the mean and median times to complete code review on the stable branch are three times shorter than those on the master and other branches. This indicates that developers prioritize completing MRs on the stable branch, which is related to the publication of new releases and may indicate the presence of bugs that could impact the delivery of the new version.

**As shown in Table 4, the stable branch receives the first comment on MRs more quickly, with mean times that are 2.81 and 2.01 times faster than those of the master and other branches, respectively.** This is further illustrated in Figure 12, which shows the distribution of time to first comment, where the stable branch consistently has shorter times compared to the

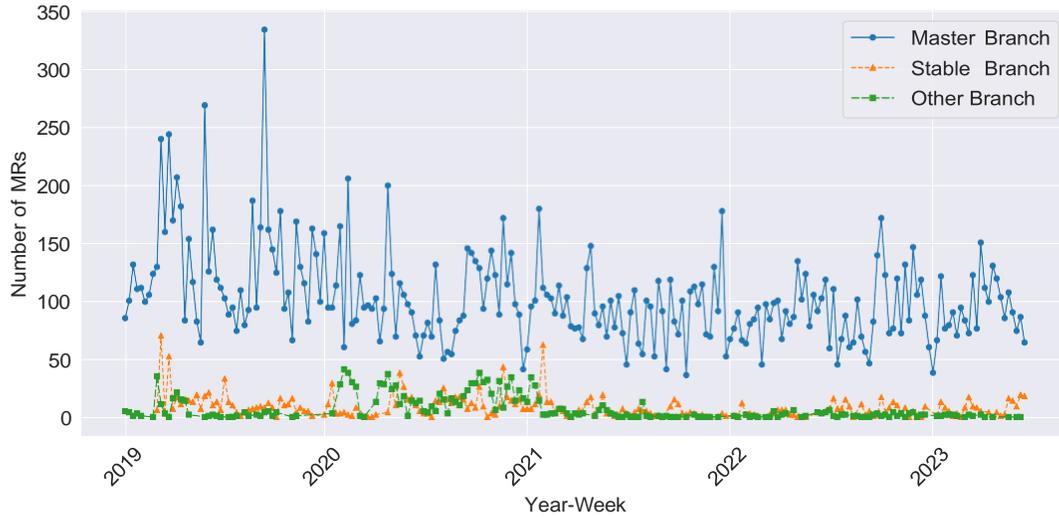

Figure 9: Weekly MRs for Master, Stable, and Other branches

other branches. Additionally, Figure 13 displays the distribution of the number of notes (comments) per MR. For the master and other branches, the largest distributions have 2 notes (23.5% and 20.8%, respectively), which then decrease, indicating fewer MRs with more notes. In contrast, the stable branch shows a different pattern, with the largest group (18.3%) having 5 notes, followed by 12.4% of MRs having 6 notes. This indicates that MRs on the stable branch are more frequently discussed than those on the master and other branches. These observations confirm that MRs on the stable branch receive more attention from team members, who respond quickly and engage in more thorough discussions. This increased attention likely helps to prevent issues or bugs during the delivery process.

| Branch | Length | Time to complete code review Mean | Time to complete code review Median | First Comment Mean | First Comment Median |
|---|---|---|---|---|---|
| master | 23,574 | 6,606.12 | 508.36 | 1,391.01 | 12.43 |
| stable | 1,923 | 2,988.93 | 118.62 | 494.79 | 0.16 |
| other | 1,467 | 6,481.96 | 227.75 | 997.07 | 10.94 |

Table 4: Statistics for Master, Stable, and Other branches

> *RQ3 summary:* MRs on stable branches are completed more quickly and receive responses faster than those on the master and other branches. This increased attention is likely due to their relation to release production, which may reflect the presence of bugs that could impact the software delivery process.

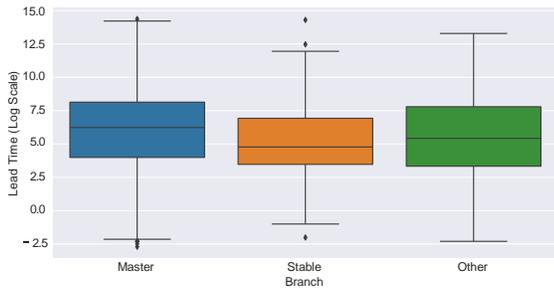

Figure 10: Time to complete code review distribution across branches

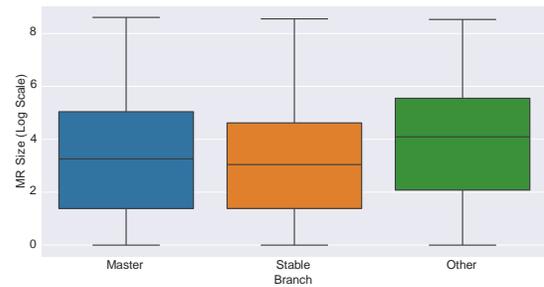

Figure 11: MR Size Distribution across Branches

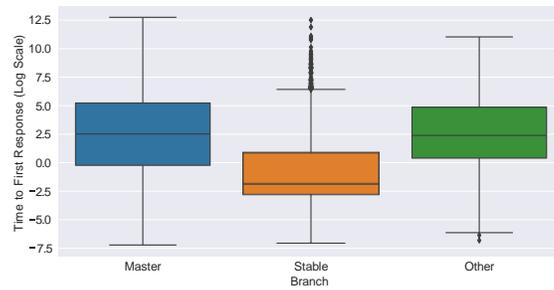

Figure 12: Time to first comment distribution across branches

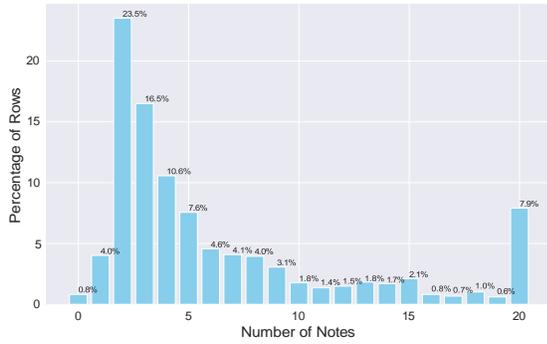

(a) Other

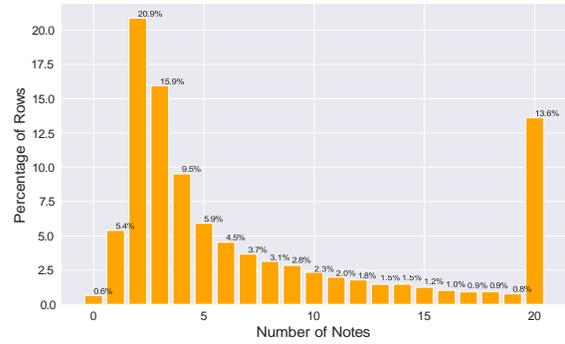

(b) Master

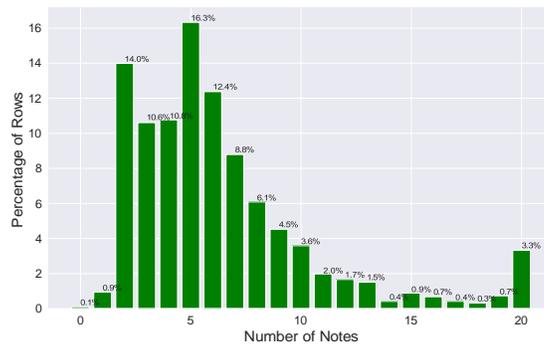

(c) Stable

Figure 13: Distribution of notes size by branch

**RQ4: How do the timing and type of initial feedback in code reviews impact mr efficiency in devops, and what roles do bots and human interactions play?**

*Motivation*

Beyond the analyses conducted in the previous RQs, code review data has been extensively studied to understand the code review process itself. One key metric is that has been widely studied is the code review completion time, which reflects the effort spent by reviewers during the integration of new code changes. Zhang et al. found that in open source context, a pull request, a mechanism very close to MR, with a shorter time to first comment is closed faster. However, their study did not differentiate between comments made by bots and those made by humans. Hasan et al. hypothesized that since bots respond quickly, the time to first comment has a weak impact on the time to complete code review of pull requests. In this research question, similar to Hasan et al. [7], we aim to analyze the code review completion time using MR data within an industrial context, with a specific focus on the time to receive the first comment, whether from a human or a bot, as it represents the initial interaction in the review process.

*Approach*

To address the research question (RQ), we followed these steps:

*Data Preparation*

- **Data collection:** We gathered data for each MR, including the discussions surrounding each MR. The collected metrics, indicated by a star (⋆), are detailed in Table 1.

- **Label First Comments:** In GitLab, each comment has a "System" parameter indicating whether it was made by a bot or not. We ran an algorithm through the MR data to determine the time between the first comment and the creation of the MR, and label each comment accordingly.

- **Define Comment Categories:** Unlike Hasan et al. [7], who used the BoDeGha tool developed by Golzadeh et al. [6] to classify comments on GitHub, we leveraged the GitLab API, which directly tags system-generated comments. We used these tags to define our categories. We categorized the types of comments to distinguish between bot and human interactions, including suspicious comments (bot-labeled but human-generated). For example, comments such as "approved this MR," which are labelled as bot comments, are actually human decisions but are classified as bot messages by the system. Table 5 provide details about different the categories used in this study.

*Data Analysis*

- **Time to First Comment Speed:** Analyze the time to receive the first comment compared to the overall time to complete the code review. This step considers the first comment, whether by a human or a bot.

- **Bot vs. Human Comment Comparison:** Compare the speediness of having the time-to-first-human-comment, time-to-first-extended-human-comment, and time-to-first-bot-comment for different types of each project of the studied company.

- **time to complete code review Analysis:** Explain the time to complete the review process on GitLab involves the following:

  – Construct a regression model to examine the correlation between collected metrics explained in the methodology section and used by Hasan et al. [7] (including time to first comment by bot and uman metrics) and MR completion time (time to complete code review). In this step, we train a mixed-effects linear regression model using the MR with comment, similarly to Zhang et al. [28].
  *Note: We train one global model using the four projects of our industrial partner, for space reason while they give all the same results*

  – Conduct an ANOVA Type-II analysis to identify statistically significant (Chisq[5] = 0.05) features that explain the variance in MR completion time, similarly to prior studies [7, 13].

  – Study the impact of key metrics on time to complete code review using Accumulated Local Effects (ALE) plots, similarly to the approach taken by Khatoonabadi et al. [12]. A negative impact indicates shorter time to complete code review probability, while positive implies longer time to complete code review probability

*Results*

**Time to first comment analysis across projects shows generally fast response times, with median values ranging from 4.46 to 23.78 minutes, as shown in Table 6.** Notably, the FPGA group shows the slowest response times, possibly due to hardware compatibility considerations impacting the interactions on GitLab. This observation suggests the need for deeper investigation into the specific hardware factors impacting response dynamics within this group. Additionally, the median time to complete code review confirms this finding, showing a significant increase for the FPGA group, exceeding 2600 minutes compared to less than 1200 minutes for the other groups. In

---
[5]Chi-square statistic: measures how well the independent variables explain variability in the dependent variable

Table 5: Definition of used categories in the study

| Term | Definition |
|---|---|
| Comment | A comment in a MR chat in Gitlab |
| Human response | A comment given in a MR by a human ( but not the MR author) |
| Bot comment | A comment given in a MR by a bot |
| Bot but human comment | A suspicious comment, written by a bot but attributable to a human |
| Human-first MR | A MR whose the first comment is given by a human (not the author of the MR) |
| Bot-first MR | A MR whose the first comment is given by a bot |
| Bot-but-human-first MR | A MR whose the first comment is a a Bot but human comment |
| Time to first comment | The time in minute from the MR creation to the first comment |
| Time to first human comment | The time in minute from the MR creation to the first human comment |
| Time to first bot comment | The time in minute from the MR creation to the first bot comment |

Table 6: Statistical analysis of the groups. Time-to-first-comment (TFC) is in minutes Avg = average, Mdn = median, Std = standard deviation.

| Project | time to complete code review | | | TFC | | |
|---|---|---|---|---|---|---|
| | Avg | Mdn | Std | Avg | Mdn | Std |
| mp | 3222.95 | 312.11 | 17032.33 | 717.32 | 11.44 | 3865.66 |
| cp | 5304.06 | 5304.06 | 37724.09 | 980.60 | 9.90 | 6132.76 |
| dp | 7690.85 | 1110.44 | 49330.31 | 1992.88 | 10.51 | 9096.11 |
| fpga | 20505.23 | 2674.39 | 54782.82 | 3992.92 | 23.30 | 13955.30 |
| pf | 6290.53 | 202.02 | 31095.99 | 1102.52 | 4.46 | 8470.01 |

other projects, the median time to first comment is faster than 11.44 minutes. For example, in the CP group, the median time to first comment is 9.90 minutes, while the median time to complete code review is 5304.06 minutes, indicating quick initial feedback compared to the total time required to complete MRs. However, the mean time and standard deviation, which reach 1992.88 and 13955.30 minutes respectively, indicate the presence of outliers that can delay the first comment.

**We observe that bot-generated comments dominate the initial interaction in the MR process and are faster compared to human comments.** As shown in Table 7, up to 69.77% of the initial comments are attributed to bots, with an additional 39.88% of comments initially made by humans but marked as bot. In contrast, human-authored comments are a minority, ranging from 7.56% to 17.58% across the projects studied. Regarding time to have the first comment, analysing Table 9, we observe that humans take significantly more time to post a comment than bots. While the median time to first human comment for all groups ranges from 212.49 minutes for MP project to 1611.87 minutes for FPGA project, the global time to first comment ranges from 4.45 minutes for PF to 23.78 minutes for FPGA, which is posted by bots. We observe the same pattern when analysing the average and the standard deviation. This comparison highlights the efficiency of the bots in initiating code reviews in a timely manner, and highlight their prevalence in accelerating the development workflow and team integration.

**As shown in Table 8, the time to first human response is the most influential, explain-**

Table 7: Proportion of bot-first MR vs human-extended-first MR

| Project | % of true-bot-first Mrs | % of bot-but-human-first Mrs | % of true-human-first Mrs |
|---|---|---|---|
| mp | 42.73% | 39.70% | 17.58% |
| cp | 59.21% | 31.80% | 8.99% |
| dp | 66.61% | 18.22% | 15.17% |
| fpga | 69.77% | 17.17% | 13.06% |
| pf | 52.55% | 39.88% | 7.56% |

Table 8: Statistical analysis of features impacting significantly the time to complete code review of the MR of our industrial partner data

| Feature | Pr(¿Chisq) | Percent |
|---|---|---|
| description_length | 5.326583e-03 | 0.2392008 |
| prev_mergereqs | 3.103913e-04 | 0.4006646 |
| is_author_final_reviewer | 9.726386e-06 | 0.6026754 |
| time_to_first_humain_response | 0.000000e+00 | 93.9714120 |
| time_to_first_bot_response | 4.705515e-02 | 0.1214756 |
| num_commits_open | 8.465914e-35 | 4.6645716 |

ing 93.97% of the variance in time to complete code review, indicating that faster human intervention to review the changes can statistically impact the overall time to complete code review of an MR. The number of commits open at the time of the MR, explaining 4.66% of the variance, also plays a crucial role in determining time to complete code review, likely due to the increased complexity and larger size of the changes, resulting in more review requirements. When the commenter is the author of the MR, the variance of the time to complete code review is impacted by 0.60%, suggesting that when authors review their own work, the process is accelerated. The number of previous MRs contributes 0.40% to the time to complete code review variance, reflecting the importance of team experience in expediting reviews. Description length contributes 0.24% of the variance, highlighting the importance of clear documentation. In addition, time to first bot response, while less influential (0.12%), shows that early bot interactions can help initiate the review process. Our interpretation is confirmed by the ALE findings indicated in Figure 14. Overall, these findings highlight that both human and automated responses, along with detailed documentation and reviewer experience, are critical to managing the efficiency of the code review process.

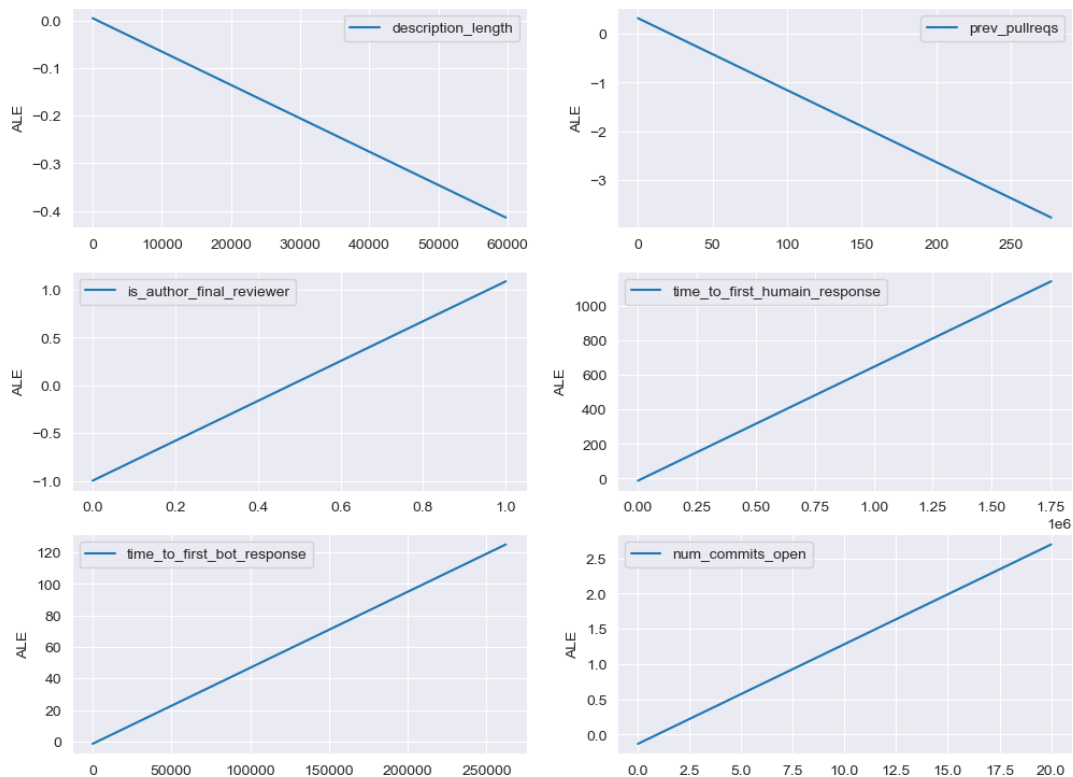

Figure 14: Impact Of the most important features on the time to complete code review

Table 9: Statistical analysis of the groups. Time-to-first-human-comment (TFHC), time-to-first-bot-but-human-comment (TFBHC), and time-to-first-human-extended-comment (TFHEC) are given in minutes.

| Project | TFHC | | | TFBHC | | | TFHEC | | |
|---|---|---|---|---|---|---|---|---|---|
| | Avg | Mdn | Std | Avg | Mdn | Std | Avg | Mdn | Std |
| mp | 2119.36 | 212.49 | 6316.11 | 690.25 | 27.81 | 3084.69 | 1098.49 | 42.75 | 4342.47 |
| cp | 4332.32 | 898.17 | 16319.31 | 1491.97 | 52.14 | 9686.19 | 2124.11 | 78.49 | 11627.50 |
| dp | 6820.87 | 1357.65 | 48153.05 | 2123.68 | 102.08 | 10758.76 | 3865.75 | 338.37 | 31395.66 |
| fpga | 14111.06 | 1611.87 | 40686.73 | 6381.94 | 166.59 | 19846.10 | 9113.04 | 921.73 | 30288.99 |
| pf | 4255.92 | 234.71 | 13388.22 | 1161.34 | 32.30 | 5099.09 | 1741.74 | 40.65 | 7810.81 |

> ***RQ4 summary:*** Bot-generated first comments dominate initial MR interactions, with median times to first comment ranging from 4.46 to 23.78 minutes. While human first comments are critical, explaining 93.97% of the variance in time to complete code review, the number of commits and reviewer experience also significantly impact time to complete code review.

## 5. Threats to Validity

**Internal Validity Threats:** These threats are related to how we calculated our metrics. We ensured that our metrics are defined according to established literature and that our implementation scripts are accurate. However, despite these precautions, there is always a possibility of implementation errors. We took steps to verify the correctness of our data and the accuracy of our scripts, but we cannot completely rule out the occurrence of errors.

**External Validity Threats:** These threats concern the generalizability of our results. Our conclusions are based on data from a single company, and work conditions can vary significantly between different organizations. Additionally, the metrics we used might need to be extended or adjusted with additional metrics to better analyze data in different contexts. The use of data from other mechanisms, such as Gerrit or pull requests, might also influence the results, as these systems have different characteristics and processes compared to GitLab.

## 6. Conclusion

This study underscores the pivotal role of MR data in offering comprehensive insights into multiple dimensions of the DevOps process, extending well beyond traditional code review analysis. By systematically examining MR data, we have gained a nuanced understanding of various aspects such as team collaboration, the impact of environmental and process changes, and the overall efficiency of the software development lifecycle.

Through the analysis of 26.7k MRs from a data center networking software company, our findings revealed that environmental changes, specifically the COVID-19 pandemic, led to a temporary increase in effort and a lasting shift in work patterns. These changes resulted in new habits that allowed greater flexibility with home working hours, with up to 70% of activities occurring outside regular office hours. Remarkably, despite these disruptions, productivity remained stable in terms of the number of MRs created and closed during the pandemic, highlighting the teams' adaptability and resilience.

The study also explored the impact of process changes, notably the migration to OpenShift technology. Initially, this transition caused fluctuations in times to complete code review and the prioritization of OpenShift-related tasks, which affected other development activities. However, as developers gained familiarity with the new technology, these metrics stabilized, reflecting successful integration and improved competency in using OpenShift.

Additionally, our research found that MRs on stable branches are consistently prioritized and resolved faster, emphasizing their significance in managing releases and addressing critical bug fixes. This prioritization illustrates the importance of efficient branch management strategies in maintaining a streamlined and reliable software delivery pipeline.

Our analysis of the code review process revealed that while automated tools (bots) play a crucial role in accelerating the initiation of code reviews by providing immediate feedback, it is the human reviewers who have the most substantial impact on reducing times to complete code review and enhancing the overall quality of code assessments. Human interactions were found to be essential in driving thorough code evaluations, with factors such as the experience of reviewers and the number of commits significantly influencing code review efficiency. This finding highlights the complementary roles of bots in expediting the review process and human reviewers in ensuring the depth and accuracy of the feedback provided.

Our findings offer practical guidance for practitioners using DevOps, illustrating how MR data can be leveraged to improve various facets of the DevOps process through detailed analysis. By exploiting MR data, organizations can enhance productivity, ensure better software quality, and adapt more effectively to both internal and external changes.